\documentclass[aps,prl,twocolumn,groupedaddress,showpacs]{revtex4}

\usepackage{graphicx}

\newcommand{\beq}{\begin{equation}}
\newcommand{\eeq}{\end{equation}}
\newcommand{\change}[1]{#1}

\begin{document}
\title{
Determining Energy Barriers by Iterated Optimization: The 
Two-Dimensional Ising Spin Glass.}

\author{C. Amoruso}
\affiliation{Institut f\"ur Theoretische Physik, Universit\"at G\"ottingen,
Tammannstrasse 1, 37077 G\"{o}ttingen, Germany }
\author{A.K. Hartmann}
\affiliation{Institut f\"ur Theoretische Physik, Universit\"at G\"ottingen,
Tammannstrasse 1, 37077 G\"{o}ttingen, Germany }
\author{M.A. Moore}
\affiliation{School of Physics and Astronomy, 
University of Manchester, Manchester M13 9PL, U.K.}

\begin{abstract}
Energy barriers determine the dynamics in many physical systems like
structural glasses, disordered spin systems or proteins.
Here we present an approach, which is based on
subdividing the configuration space in a hierarchical manner, leading to
upper and lower bounds for the energy barrier separating two
configurations. The fundamental operation is to perform
a constrained energy optimization, where the degree of
constraintness increases with the level in the hierarchy.

As application, we consider Ising spin glasses, where
the energy barrier which needs to be surmounted in order to flip a
compact region of  spins of linear dimension $L$ are expected
to scale as $L^{\psi}$. The exponent $\psi$ is
very hard to estimate from experimental and simulation studies.
By using the new approach, applying efficient combinatorial 
matching algorithms,  we are able to give the first non-trivial numerical  
bounds  $0.25 < \psi < 0.54$ for the  two-dimensional  Ising spin glass.
%{\bf Keywords (PACS-codes)}: Spin glasses and other random models (75.10.Nr), 
%numerical simulation studies (75.40.Mg).
 %General mathematical systems (02.10.Jf).
\end{abstract}
\maketitle

A unifying concept in the physics of
disordered systems is the notion of energy landscapes.
\change{The dynamics of these systems is determined by (free) energy
  barriers.
Prominent examples  include spin glasses \cite{reviewSG}, structural
glasses \cite{reviewGlass} and folding of
proteins \cite{creighton1992}. The barrier problem is
also related to theoretical computer science, 
because it belongs \cite{middleton1999} to the
fundamental class of nondeterministic polynomial (NP) hard 
problems \cite{garey1979}, hence all known
algorithms determining lowest barriers take a time
growing exponentially with system size.

In this paper, we will present a hierarchical approach to calculate
minimum barriers, which is based on the application of combinatorial
optimization algorithms
\cite{opt-phys2001}. As application, we apply it to the
prototypical two-dimensional Ising spin glass, where the value of the
barrier-height exponent could not be determined so far.}

In the droplet \cite{FH1} or scaling theory of Ising spin glasses the
low-energy excitations  are compact droplets. The creation of a
droplet (a region of reversed spins) results in the formation of a
domain wall around the reversed spins and the energy of the droplet
scales as $L^\theta$ where $L$ is the linear extension of the droplet.
The system orders at low temperatures only if $\theta > 0$. The
domain wall is  fractal and has \change{size}
  $L^{d_S}$, with $d-1\leq d_S \leq d$.   The dynamics of the
system are controlled by the height of the barriers which have to be
crossed to create such droplets. It is generally assumed that the
barrier to be surmounted to create a droplet of linear extent $L$ has
an energy  which scales as $L^\psi$ where $\theta \leq \psi \leq d-1$
for dimension $d$.  The argument for these inequalities goes as
follow: the barrier must be at least as large as the energy required
to create the domain, hence $\theta \leq \psi$; the upper limit is due
to the fact that the barrier must be lower  than the energy of a
compact droplet with a non-fractal surface containing  the same number
of reversed spins.  While for a directed polymer in a random system it
was shown that  $\psi = \theta$, for Ising spin glasses $\psi$ appears
to be an independent exponent.

While many numerical studies have been done to calculate the exponent
$\theta$, we are aware of only one direct numerical estimate of $\psi$
in two dimension \cite{GCB}. It was equal to the upper bound  i.e,
$\psi = d-1$, but  only small systems were studied ($L \leq 6$).
\change{As mentioned, the barrier problem is NP hard and} it has been
difficult even to find good approximate algorithms for barriers.  This
is presumably the reason for the paucity of studies of this exponent.
Recently Drossel and Moore instead of  attempting to calculate the
barrier exactly,  placed bounds on its energy \cite{MD}. They showed
that  for the hierarchical (i.e. Berker) lattice there exists a lower
bound on $\psi$  which is the same as the upper bound $d-1$ and hence
they concluded that $\psi = d-1$.  In the same spirit as their work,
we will also obtain upper and lower bounds on the energy of the
barriers but for the more physically relevant case of the square
lattice.  Rather than study droplets, we will consider the
computationally simpler (but equivalent) task of determining  upper
and lower bounds for  the  barrier \change{separating the 
two ground states (GSs) 
related by a flip of all spins. The algorithm is based on
subdividing the configuration space in a hierarchical manner and
performing constrained energy optimizations, where the degree of
constraintness increases with the level in the hierarchy. This
approach can be applied in general to obtain minimum barrier for many 
problems, but for
the convenience of the reader, we formulate it here for the case of
the two-dimensional Ising spin glass.}

The model consists of $N=L^2$ spins $S_i = \pm 1$ on a  
square lattice with periodic boundary conditions in the $x$ direction
and free boundary conditions in the $y$ direction.
The Hamiltonian  is
\beq
H = -\sum_{\langle i j \rangle}J_{ij}S_i S_j,
\eeq
where the sum runs over all pairs of nearest 
neighbors $\langle i j \rangle$ and the $J_{ij}$ are the
quenched random spin couplings. We will consider a Gaussian
distribution of couplings with zero mean and unit width.
We define the barrier to be the energy required to invert all spins
in the GS within the context of single spin flip dynamics,
as in \cite{GCB}. Each trajectory of reversal is characterized by  
the highest maximum in the trajectory having
energy  $\delta E_{max}$, in excess of the GS energy, so that
$E_B = \min(\delta E_{max})$, where the minimum has to be computed over
all the possible trajectories.

Thus the first task is to compute the GS $\{\sigma^0\}$
of the spin glass;
in general already this problem is NP hard, but for planar graphs 
it can be mapped on to  the {\em minimum-weight perfect matching} problem,
which is solvable in polynomial time. 
The algorithms for minimum-weight perfect matchings 
\cite{MATCH-cook,MATCH-korte2000} are
among the most complicated algorithms for polynomial problems but
 the LEDA library offers a very efficient implementation
\cite{PRA-leda1999}.
For the details of the method, see
Refs.\ \onlinecite{opt-phys2001,matching-all}. 

We will now show how is it possible to extract bounds on $E_B$.
\change{In general, a lower bound on the energy barriers can be
obtained, if one considers a sequence of constraints fixing some
degrees of freedeom leading from one
configuration to the other and calculating the GS under each
constraint. Here, we force a domain wall, which seperates spins having
orientation $\{\sigma^0\}$ from spin having orientation $\{-\sigma^0\}$, 
through constraining}
successive bonds along the x-axis.  Thus, for each realization of the
bonds $J_{ij}$,  a GS is first calculated with free boundary
conditions in both directions. Then ``hard'' bonds \cite{droplets}
are introduced,
i.e. bonds with a high value of the absolute strength, compatible with
the relative GS orientation of the adjacent spins, in a line
which runs from the left border to the right border. Next the sign of
exactly one hard bond on this line is inverted and the GS of
the new realization is calculated.  \change{$L$ times exactly one bond is
inverted.} The minimum of these $L$ energies
is the domain wall associated with the exponent $\theta$. The maximum
can be associate with a lower bound on $\psi$, because one is
calculating  the minimum with the highest energy in the sequence from
putting the hard bonds at successive places on the x-axis. The true
barrier height is associated with flipping one spin at a time and one
might have to climb an even higher barrier to reach the maximum domain
wall energy starting from the domain wall of minimum energy. We will
next explain how to place an upper bound on the barrier height before
discussing  the results of the numerical work.

\begin{figure}[htb] 
\includegraphics[angle=0,width=0.80\columnwidth]{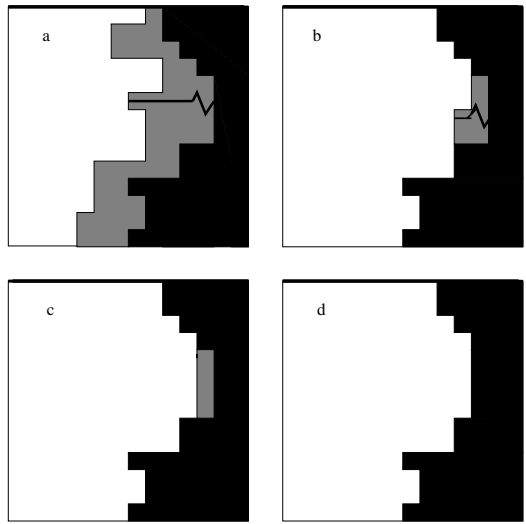}
\caption{Example showing how  the domain wall is shifted. In (a) the black
  and the gray area represent the domain walls created by changing the
  sign of two consecutive  hard bonds in the top (thick) line. The
  maximum distance between the two domain walls is determined and hard
  bonds are introduced along this line (thick line at the
  middle). Then the sign of each of these bonds is inverted exactly
  once (wiggled segment) and the new GS is calculated,
so that the  domain wall is forced   to pass through them .
In (b) the grey area now represents the configuration obtained when
changing the sign of the leftmost hard bond and calculating the new
GS. Again hard bonds are introduced along the line of
maximum distance and the new GS (c) is computed for the
realization  in which the leftmost hard bond is
inverted, as in (a). Now the distance between the two domain walls is at most one spin,
so this spin is now flipped (yielding one step of the whole
sequence) and the resulting configuration is shown in (d). 
}
\label{fig:matching}
\end{figure}

\change{The basic idea of the general algorithm is to continue the constrained
minimization in a hierarchical manner, i.e. one introduces additional
constraints between all adjacent pairs of constrained GSs
obtained above. This is continued recursively, until the
level of changes of single degrees is reached. Hence, one obtains a
path in configuration space where the maximum energy configuration 
yields an upper bound for the  minimum energy
barrier. I.e. For the spin glass
we} do this by introducing  hard bonds along the
line corresponding to the maximum spacing between two ``adjacent'' domain
walls (either in the $x$ or the $y$ direction) and by forcing the
first domain wall through each of these bonds, see Fig.\
\ref{fig:matching}. This procedure is
repeated iteratively until two consecutive domain walls differ only by
a unique sequence of  spin flips.
Of course, since we are minimizing over a restricted set of paths,
there could be some lower path over the energy landscape, hence our
result will be an upper bound on the energy barrier. However, since
the upper bound makes an attempt to estimate the energy needed to move
the domain wall to successive places along the  x-axis by a sequence
of single spin flips, we believe the true value of $\psi$ is likely to
be nearer the upper bound than the lower bound.

\begin{figure}[htb] 
  \includegraphics[angle=0,width=0.80\columnwidth]{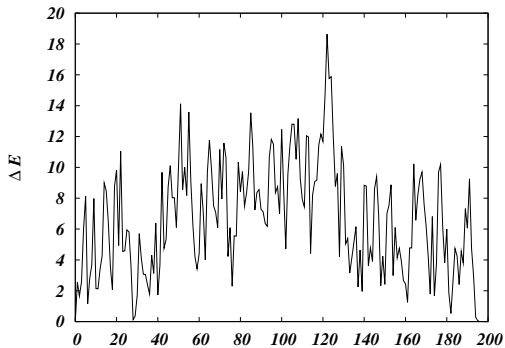}
  \caption{An example of the energy landscape explored by our
    algorithm for $L=14$; the exponent $\psi$ is  associated with the scaling of the maximum energy found in the sequence.
  }
  \label{Landscape.fig}
\end{figure}

For each flipped spin the energy of the configuration is
calculated, and at the end the maximum among the $N$ energies is
considered. In this way we give an explicit rule to build a sequence
of spin-flip taking one from the GS configuration $\{\sigma^0\}$ to
$\{-\sigma^0\}$.
An example of the corresponding energy landscape sampled by this
sequence is shown in Fig. \ref{Landscape.fig}.\\
One has to be careful to exclude the  domain walls
which are not crossing the system and for which the width $L$
is not an appropriate measure of their scale. Those configurations can appear in the first
steps of the algorithm described above because of the use of free
boundary conditions. 
Fortunately most of the domain walls do span the system. We have restricted our analysis to
 those configurations.
\begin{figure}[htb] 
  \includegraphics[angle=0,width=0.80\columnwidth]{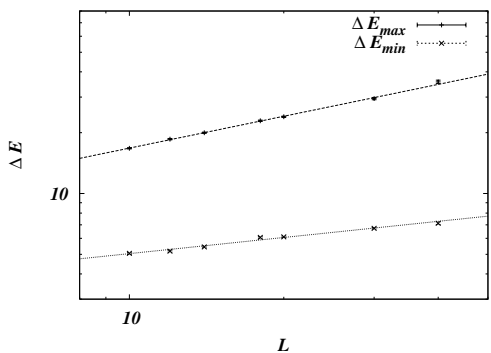}
  \caption{Upper line: maximum over $L$ energies in the sequence from 
    putting hard bonds at successive places on the $x$ axis  plotted  as a function of the system size.
    Lower line: maximum energy in the sequence of single spin-flip
    built as explained above.
  }
  \label{fig:barriers}
\end{figure}

We applied our method for system sizes in the range $8 < L < 40$,
and for each $L$ used  $1000$ independent realizations of the bonds $J_{ij}$.
The two quantities $\Delta
E_{min}$ (the lower bound) and $\Delta E_{max}$ (the upper bound) as a function of system size are shown
in Fig. \ref{fig:barriers}. Both quantities can  be fitted by an
algebraic function $\sim L^{\psi}$, where $\psi_{lw} = 0.25 \pm 0.01$
and $\psi_{up} =  0.54 \pm0.01$. 
There could of course be a
procedure to obtain a better path in
configuration space, yielding a barrier which grows more slowly than
$L^{0.54}$. However this is to our knowledge the only numerical estimate of
the energy barriers exponent giving an upper bound different from the
trivial one $d-1$.

There are many experimental estimate of $\psi$ in the literature, and
there is disagreement as to its value.  Dekker et al \cite{EXP1}
reported experimental verification of activated dynamics in a $d=2$
system obtaining $\psi = 0.9$.  Schins et al. \cite{EXP2} find $\psi =
1.0 \pm 0.1$ by studying aging via the low-frequency ac susceptibility
giving credence to the claim that $\psi$ is equal to its upper bound.
However, Dupius et al \cite{EXP3} extracted $0.3 < \psi < 0.7$ in
$d=2$ and $\psi \sim 1.1$ in $d=3$.  The numerical study of Gawron et
al. gave  $\psi = 0.9 \pm 0.1$ in 2 dimension, while that of Berthier
and Bouchaud \cite{BerthBouch3d} had $\psi \sim 1.0$ and $\psi \sim
2.3$ in 3 and 4 dimensions respectively.  We remark that our
optimization algorithms do not suffer from equilibration problems,
they are exact and they allow the study of large systems at least in
two dimensions. Besides, by iteratively applying our matching
algorithm, we are producing large scale changes to the domain wall,
which is how $\psi$ can become less than $d-1$. (We have studied the
barrier associated with a straight domain wall and found that the
expected linear behavior is approached  by $L\sim 40$).

\begin{figure}[t] 
  \includegraphics[angle=0,width=0.75\columnwidth]{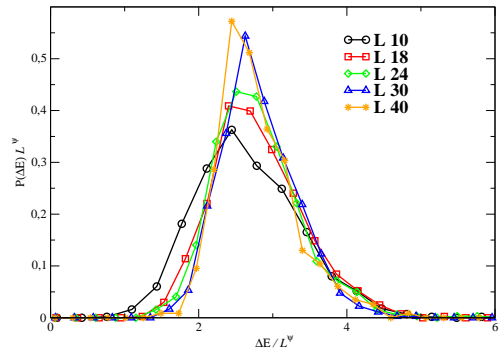}
  \caption{Rescaled probability distribution of the lower bound for the
barriers heights with $\psi = 0.25$}
  \label{distrib_LB.fig}
\end{figure}

\begin{figure}[htb] 
  \includegraphics[angle=0,width=0.80\columnwidth]{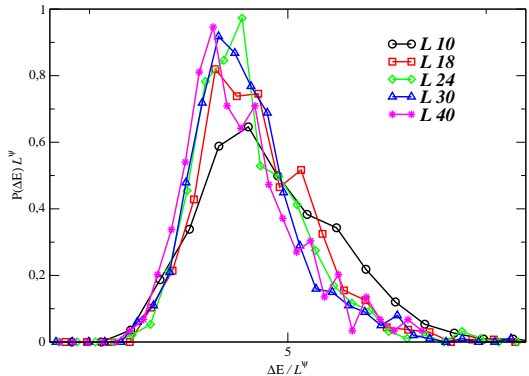}
  \caption{Rescaled probability distribution of the upper bound for the
barriers heights corresponding to $\psi = 0.54$}
  \label{distrib_UB.fig}
\end{figure}

In Figs.\ \ref{distrib_LB.fig} and \ref{distrib_UB.fig}  
the rescaled probability
distributions of the barriers heights corresponding to both the lower
and upper bounds which we have found,  averaged over the disorder, is
shown for different sizes $L$: in both cases it seems to approach  a
fixed shape in the thermodynamical limit (although the approach to a
scaling collapse is faster for the upper bound).


Notice that both these distributions have long tails indicating the
existence of some very large barriers, 
which implies that some spin configurations could be very long-lived.
%

%
%
%Besides one can show that if the lower bound  follows a power law
%scaling, then the same holds for tail of the energy barrier
%distribution. Consider in fact the probability distribution of domain
%walls passing through each bond in turn for a given bond realization,
%$P_L(E)$. The probability that all the $L$ values are less than $E$ is 
%$Q(E) = [\int_0^E dJ P(J)]^L \simeq exp^{-L\int_E^{\infty} dJ P(J)}$,
%so that $P_L(E) = \frac{dQ}{dE} = L Q(E)P(E)$. Now if we suppose that,
%for large $J$,$P(J) \sim J^{-(1+\alpha)}$, then $P_L(E)$ has a peak $E
%\sim L^{-\alpha} = L^{\psi_{LB}}$. This means the $\alpha =
%\frac{1}{\alpha} \simeq 4$.

We also investigated geometrical properties of the domain walls
corresponding to the energy barriers.
The boundaries of the low-energy excitations are fractal with a
an average length  along the perimeter of order  $P \sim L^{d_S}$,
with $d-1 \leq d_S \leq d$. 
\begin{figure}[htb] 
  \includegraphics[angle=0,width=0.80\columnwidth]{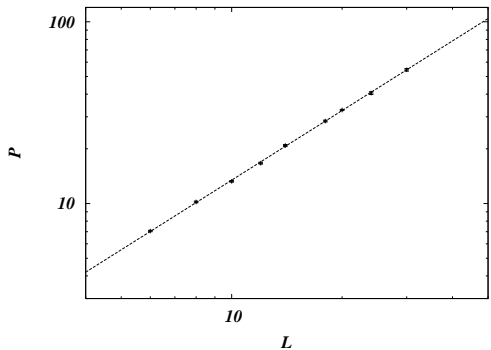}
  \caption{The perimeter (length) of the surface of the domain wall corresponding to the upper bound (maximum)
energy.
 Data are fitted very well by
    a power law $P \sim L^{d_S}$ with $d_S = 1.27 \pm 0.01$.}
  \label{fig:perimeter}
\end{figure}
We have computed the fractal dimension of the domain wall $d_S$ corresponding
to the configuration of maximum energy. We have found the
interesting result that $d_S = 1.27 \pm 0.01$, which is the same value 
 for domain walls of minimum energy.

%This feature
%suggests the following speculation: in the droplet theory, the
%smallness of  $L^{\theta}$ compared to the {\em naive } estimate
%$L^{d-1}$ for the free energy of an overturned region of spins of size
%$L$ is due to large cancellation from the energetic term $\sim L^{\frac{d_s}{2}}$
%and the entropic term $\sim \sigma(T)L^{\frac{d_s}{2}}$. At zero
%temperature this is no more true and since  the
%carachteristic lenght of the configuration  corresponding to the maximum
%energy is the same as the one corresponding to the minimum energy; 
%this could imply  $\psi = \frac{d_s}{2}$,  which is consistent
%with our estimation of $\psi$ and $d_s$.

%**************************************************************************

%**************************************************************************
To conclude we \change{have introduced a hierarchical algorithm to compute
upper and lower bounds on energy barriers in disordered
system. Applying the algorithm to two-dimensional Ising spin glasses,
we find that the minimum barrier energy}
is bounded above by the scale
$L^{0.54}$ and below by $L^{0.24}$. Our numerical upper bound is
significantly less than the rigorous upper bound (and the value for
the hierarchical lattice), $d-1$. Hence, we suspect that 
$\psi$ is an  entirely non-trivial exponent. At the present time there
seems no theoretical approach which might explain a non-trivial value
of $\psi$

%**************************************************************************

%**************************************************************************
The authors have obtained financial support from the
{\em VolkswagenStiftung} (Germany) within the program
``Nachwuchsgruppen an Universit\"aten'',
\change{from the Paderborn Center
for Parallel Computing in German,
 and from the European Community
via the DYGLAGEMEM, High-Level Scientific
Conferences (HLSC) and the  ``Exystence'' programs.}
%The simulations were performed at the Paderborn Center
%for Parallel Computing in Germany and on a workstation
%cluster at the Institut f\"ur Theoretische Physik, Universit\"at
%G\"ottingen, Germany.

\end{document}